\documentclass{main}

\def\be{\begin{equation}}
\def\ee{\end{equation}}
\def\bea{\begin{eqnarray}}
\def\eea{\end{eqnarray}}

\def\Pom{{\bf I\!P}}
\def\Reg{{\bf I\!R}}
\newcommand{\Photo}{}

\begin{document}

\title{\Large 
Light-by-light scattering in ultraperipheral collisions \\
of heavy ions with future FoCal and ALICE 3 detectors}

\author{Antoni Szczurek}

\address{Institute of Nuclear Physics, PAN, Krak\'ow \\
ul Radzikowskiego 152, PL-31-342 Krak\'ow, Poland and\\
Rzesz\'ow University \\
ul. Pigonia 1, PL-35-310 Rzesz\'ow, Poland
}

\maketitle

\abstracts{
I present possible future studies of light-by-light scattering
using FoCal@ALICE and ALICE 3 detectors.
Different mechanisms are discussed.
The PbPb$\to$PbPb$\gamma \gamma$ cross section is calculated
within equivalent photon approximation in the impact parameter space.
Several differential distributions are presented and discussed.
 We predict cross section in the (mb-b) range for typical ALICE 3 
cuts, a few orders of magnitude larger than for the current 
ATLAS or CMS experiments.
We also consider the two-$\pi^0$ background which can, in principle, 
be eliminated at the new kinematical range
for the ALICE 3 measurements by imposing dedicated cuts on diphoton 
transverse momentum and$\slash$or so-called vector asymmetry.
}


\section{Introduction}

Photon-photon scattering is purely quantal effect (does not happen
in classical physics). It was studied experimentally only recently
in ultrarelativistic heavy ion collisions by the
ATLAS \cite{ATLAS} and CMS \cite{CMS} collaborations.
The ATLAS and CMS kinematics implies that box diagrams are
the dominant reaction mechanism and other mechanisms are practically
negligible. In \cite{KNSS} we studied whether this process could be
sudied at lower photon-photon energies using ALICE and LHCb 
infrastructures. The experimental analysis using ALICE data
is in progress. At the lower energies one should worry about
background due to $\gamma \gamma \to \pi^0 \pi^0$ process.
We have worked out there techniques how to reduce the unwanted 
background \cite{KNSS}.

The $\gamma\gamma \to \gamma\gamma$ is also interesting in 
the context of searching for effects beyond Standard Model 
\cite{Baldenegro}.

Recently we explored what future FoCal \cite{FoCal} and 
ALICE 3 \cite{ALICE3} detectors could do in this respect
\cite{JKS2024}.
A forward electromagnetic calorimeter is planned as an upgrade to 
the ALICE experiment for data-taking in 2027-2029 at the LHC. 
The FoCal will cover pseudorapidities range of $3.4<\eta<5.8$. 
Runs 5 and 6 will measure more than five times the present 
Pb-Pb luminosity. This increase of luminosity, in combination with 
improved detector capabilities, will enable the success of 
the physical program planned for ALICE 3. A significant feature of 
FoCal and ALICE 3 programs is the ability to measure photons 
in relatively low transverse momenta. 

\begin{figure}[!h]
	(a)\includegraphics[scale=0.3]{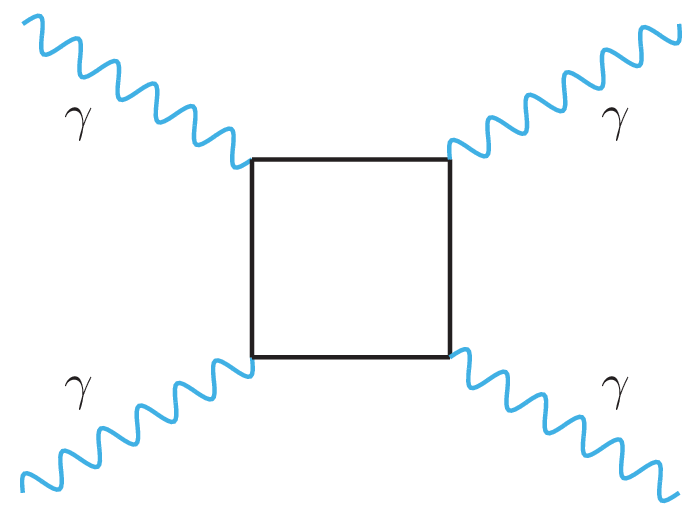}
	(b)\includegraphics[scale=0.45]{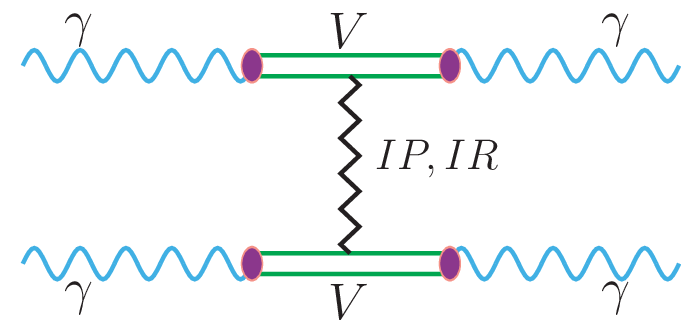}
	(c)\includegraphics[scale=0.45]{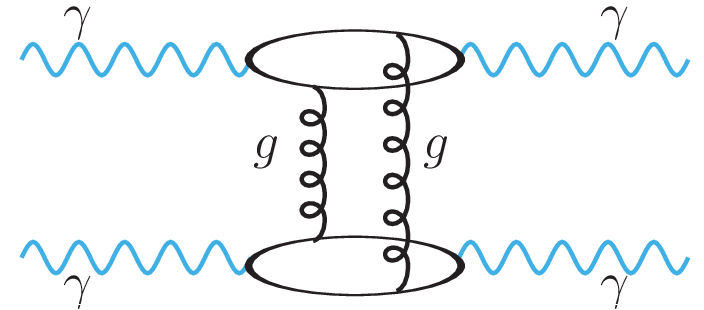}
  \begin{center}
    (d)\includegraphics[scale=0.3]{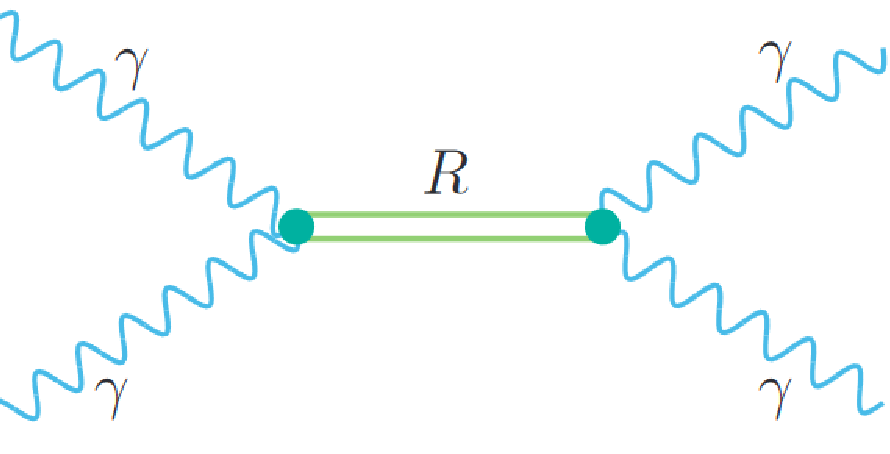} 
    (e)\includegraphics[scale=0.25]{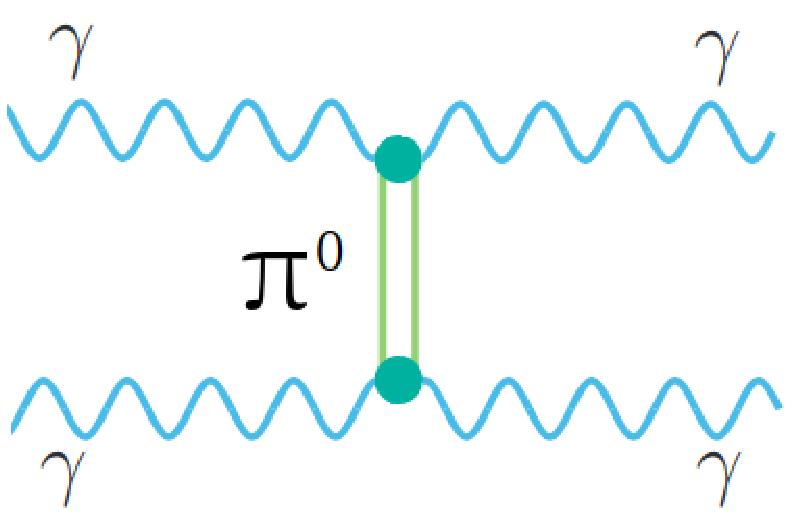}
  \end{center}
	\caption{Feynman diagrams representing different mechanisms: 
          a) fermionic loops,
          b) VDM-Regge, c)~2-gluon exchange, d) low mass resonances in
          s-channel, e) $\pi^0$-exchange in t-channel.}
	\label{fig:Feynman}
\end{figure}

In our recent paper we have taken into account different mechanisms 
for photon-photon scattering: boxes, VDM-Regge, two-gluon exchanges,
meson (resonance) contributions, see Fig.\ref{fig:Feynman}.

Here, in this presentation, I will present the main results 
from \cite{JKS2024}.

\section{\label{sec:level2}  Sketch of the formalism}

Here we briefly discuss only some selected problems discussed
in detail in \cite{JKS2024}.

\subsection{Double-photon hadronic fluctuations}

This component was calculated for the first time in \cite{KLS2016}
assuming vector dominance model. In this approach, the amplitude 
for the process is given as:
\begin{eqnarray}
{\cal M} &=& \Sigma_{i,j} C_i^2 C_j^2 \left( C_{\Pom} \left( \frac{s}{s_0} \right)^{\alpha_{\Pom}(t)-1} F(t)
                                + C_{\Reg} \left( \frac{s}{s_0} \right)^{\alpha_{\Reg}(t)-1}F(t) \right)
\; ,
\nonumber \\
         &+& \Sigma_{i,j} C_i^2 C_j^2 \left( C_{\Pom} \left( \frac{s}{s_0} \right)^{\alpha_{\Pom}(u)-1} F(u)
                                + C_{\Reg} \left( \frac{s}{s_0} \right)^{\alpha_{\Reg}(u)-1} F(u) \right)
\; .
\label{SS_amplitude}
\end{eqnarray}
In the simplest version of the model $i, j = \rho^0, \omega, \phi$ 
(only light vector mesons are included).
The couplings $C_i, C_j$ describe the $\gamma \to V_{i/j}$ transitions
that are calculated based on vector meson dilepton width.
$C_{\Pom}$ and $C_{\Reg}$ are extracted from the Regge 
factorization hypothesis.

It was shown in \cite{KLS2016} that the component is 
concentrated mainly 
at small photon transverse momenta which at not too small subsystem
energies corresponds to $z \approx \pm$ 1.
The Regge trajectories are usually written in a linear form:
\begin{eqnarray}
\alpha_{\Pom}(t/u) = \alpha_{\Pom}(0) + \alpha_{\Pom}'t/u \; , \nonumber \\
\alpha_{\Reg}(t/u) = \alpha_{\Reg}(0) + \alpha_{\Reg}'t/u \; .
\label{linear_trajectories}
\end{eqnarray}
These linear forms are valid at not too large $|t|$ or $|u|$.
At large $|t|$ or $|u|$ the energy dependent factors are artificially 
small.
In \cite{JKS2024} we proposed to smoothly
switch off the $t/u$ dependent terms in (\ref{linear_trajectories})
at $t, u \sim$ -0.5 GeV$^2$. 
The actual place where it should be done is not known precisely.
Another option would be to use $\sqrt{t/u}$ trajectories 
\cite{Brisudova2,Brisudova1}. 

We also analyzed in \cite{JKS2024} whether more heavy vector mesons 
such as $J/\psi$ can give a sizeable contribution.

For the double $J/\psi$ fluctuations 
(both photons fluctuate into virtual $J/\psi$ mesons) 
we took the following Ansatz for the helicity
conserving amplitude:
\begin{eqnarray}
{\cal M}_{VDM}^{J/\psi J/\psi} &=& g_{J/\psi}^2 C_{\Pom}^{J/\psi} 
\left( \frac{s}{s_0} \right)^{\alpha_{\Pom}^{J/\psi J/\psi}(t) - 1}
  F_{J/\psi J/\psi \Pom}^{H}(t) F_{J/\psi J/\psi \Pom}^{H}(t)  
\nonumber\\
                               &+& g_{J/\psi}^2 C_{\Pom}^{J/\psi}
\left( \frac{s}{s_0} \right)^{\alpha_{\Pom}^{J/\psi J/\psi}(u) - 1}
  F_{J/\psi J/\psi \Pom}^{H}(u) F_{J/\psi J/\psi \Pom}^{H}(u)
\; .
\label{HH_amplitude}
\end{eqnarray}
In this case (double $J/\psi$ fluctuations) only pomeron can be exchanged 
(no subleading reggeons are possible due to the pure $c \bar c$
structure of $J/\psi$ ).
In this case, for simplicity, we took the simplified trajectories as
\begin{equation}
\alpha_{\Pom}^{J/\psi J/\psi}(t) = \alpha_{\Pom}^{J/\psi J/\psi}(u) =
\alpha_{\Pom}^{J/\psi J/\psi}(0) \; .
\label{trajectory_forJpsiJpsi}
\end{equation}
Here the $t/u$ dependencies of the trajectories are totally ignored.
In numerical calculations we take $\alpha_{\Pom}^{J/\psi J/\psi}(0) =
1.3 - 1.4$ (typical hard pomeron).
Since the $J/\psi$ mesons are far off-mass-shell and more compact than light
vector mesons the form factors are different than those for light 
vector mesons. 
In \cite{JKS2024} we took them in the following form:
\begin{eqnarray}
F_{J/\psi J/\psi \Pom}^H(t) = \exp\left( \frac{t-m_{J/\psi}^2}{\Lambda_{J/\psi}^2}
\right) \; , \\
F_{J/\psi J/\Psi \Pom}^H(u) = \exp\left( \frac{u-m_{J/\psi}^2}{\Lambda_{J/\psi}^2}
\right) \; .
\label{hard_formfactors}
\end{eqnarray}
These form factors are normalized to 1 on the meson ($J/\psi$)
mass shell. One could also use monopole form factors in this context.
The form factors reduce the $J/\psi J/\psi$ component of the amplitude 
in comparison to light vector meson components.
However, due to compactness of $J/\psi$ we expect $\Lambda_{J/\psi}$ to 
be large. 
In \cite{JKS2024}we took $\Lambda_{J/\psi}=2$~GeV
for illustration, the actual value is not precisely known.
Also, the normalization parameter $C_{\Pom}^{J/\psi}$ is not well known.
It is expected to be smaller than for light vector mesons.

In a similar fashion, one could include one $J/\psi$ fluctuation and 
one light vector meson fluctuation. However, there the choice of 
trajectories is not clear. We will leave the discussion of these 
components for future studies.

In \cite{JKS2024} we assumed the following helicity structure
of the double photon hadronic fluctuation amplitude:
\begin{eqnarray}
{\cal M}_{\lambda_1 \lambda_2 \to \lambda_3 \lambda_4}^{(t)} &=&
A(t) \; \delta_{\lambda_1 \lambda_3} \delta_{\lambda_2 \lambda_4} \; , \\
{\cal M}_{\lambda_1 \lambda_2 \to \lambda_3 \lambda_4}^{(u)} &=&
A(u) \; \delta_{\lambda_1 \lambda_4} \delta_{\lambda_2 \lambda_3} \; .
\label{amplitudes}
\end{eqnarray}
$A(t)$ and $A(u)$ are given explicitly in (\ref{SS_amplitude}).
Then the total double VDM amplitude, including $t$ and $u$ processes, 
reads:
\begin{equation}
{\cal M}^{VDM}_{\lambda_1 \lambda_2 \to \lambda_3 \lambda_4} =
\frac{1}{\sqrt{2}} \left(
{\cal M}^{VDM,(t)}_{\lambda_1 \lambda_2 \to \lambda_3 \lambda_4} +
{\cal M}^{VDM,(u)}_{\lambda_1 \lambda_2 \to \lambda_3 \lambda_4}
\right)
\; .
\end{equation}
Now we can add amplitudes for different mechanisms:
\begin{equation}
{\cal M}_{\lambda_1 \lambda_2 \to \lambda_3 \lambda_4} =
{\cal M}^{boxes}_{\lambda_1 \lambda_2 \to \lambda_3 \lambda_4} +
{\cal M}^{VDM}_{\lambda_1 \lambda_2 \to \lambda_3 \lambda_4} +
{\cal M}^{\pi^0}_{\lambda_1 \lambda_2 \to \lambda_3 \lambda_4} + ... \; .
\label{summing_amplitudes}
\end{equation}
In the following, we shall discuss the sum of the larger two components 
(boxes and VDM) and quantify their interference effects.

\subsection{Cross section for nuclear UPC}
\label{subsec:nuclear_cs}

In \cite{JKS2024} the nuclear cross section is calculated using 
equivalent photon approximation (EPA) in the b-space. In this approach,
the diphoton cross section can be written as (see \cite{KLS2016}):

\begin{eqnarray}
    \frac{d\sigma(PbPb \to PbPb \gamma \gamma)}{dy_{\gamma_1}dy_{\gamma_2}dp_{t,\gamma}} &=& 
    \int \frac{d\sigma_{\gamma\gamma\to\gamma\gamma}(W_{\gamma\gamma})}{dz}N(\omega_1,b_1)N(\omega_2,b_2)S^2_{abs}(b) \nonumber \\  
    &\times& d^2b d\bar{b_x} d\bar{b_y} \frac{W_{\gamma\gamma}}{2} \frac{dW_{\gamma\gamma}dY_{\gamma\gamma}}{dy_{\gamma_1}dy_{\gamma_2}dp_{t,\gamma}} dz \;,
    \label{eq:tot_xsec}
\end{eqnarray}
where $\bar{b}_x = \left( b_{1x} + b_{2x}\right)/2$ and $\bar{b}_y =
\left( b_{1y} + b_{2y}\right)/2$. The relation between $\vec{b}_1$,
$\vec{b}_2$ and impact parameter: $b = |\vec{b}| = \sqrt{|\vec{b}_1|^2 +
  |\vec{b}_2|^2 - 2|\vec{b}_1||\vec{b}_2|\cos\phi}$. Absorption factor
$S^2_{abs}(b)$ in \cite{JKS2024}was calculated as:

\begin{eqnarray}
    S^2_{abs}(b) = \Theta(b-b_{max})  
\end{eqnarray}
%
or
\begin{eqnarray}
    S^2_{abs}(b) = exp\left( -\sigma_{NN} T_{AA}(b) \right) \;,
    \label{eq:s2b}
\end{eqnarray}
%
where $\sigma_{NN}$ is the energy-dependent nucleon-nucleon 
interaction cross section, and $T_{AA}(b)$ is related to the so-called nuclear thickness, $T_A(b)$,
\begin{equation}
    T_{AA}\left(|\vec{b}| \right) = \int d^2\rho T_A \left( \vec{\rho} -\vec{b} \right) T_A\left(\rho\right) ,
\end{equation}
%
and the nuclear thickness is obtained by integrating the nuclear density 
\begin{equation}
    T_A \left( \vec{\rho} \right) = \int \rho_A\left( \vec{r} \right) dz , \hspace{0.5cm} \vec{r} = \left( \vec{\rho},z \right) \;,
\end{equation}
%
where $\rho_A$ is the nuclear charge distribution.
The nuclear photon fluxes $N(\omega_1,b_1)$ and $N(\omega_2,b_2)$ 
are calculated using realistic charge distribution.

Very often the UPC results are shown only with 
a sharp cut on the impact parameter, usually taken as a sum 
of two radii of the nuclei, i.e. $b> R_A + R_B \approx 14$~fm 
for Pb+Pb collisions. 
Due to the no homogeneous nuclear charge distribution, it seems 
to be more reasonable to use the absorption factor given by 
Eq.~(\ref{eq:s2b}).

\subsection{Background contribution}

As shown in \cite{KNSS} the 
$\gamma \gamma \to \pi^0(\to2\gamma)\pi^0(\to2\gamma)$ reaction 
constitutes a background for the measurement 
of $\gamma\gamma \to \gamma\gamma$ process at intermediate 
$M_{\gamma\gamma}$. 

In our approach the calculation of the background proceeds 
in three steps. 
First, the cross section for $\gamma\gamma \to \pi^0\pi^0$ is 
calculated (for details, see \cite{KS2013}). 
Next the cross section for $AA \to AA\pi^0\pi^0$ is computed in 
the equivalent photon approximation in an analogous way as 
described in the subsection above. In the last step the simulation 
of both $\pi^0$ decays is performed and joint distributions 
of one photon from the first $\pi^0$ and one photon from 
the second $\pi^0$ are constructed.

\section{Selected results}

\subsection{Elementary cross section}

\begin{figure}[!h]
        \begin{center}
	(a)\includegraphics[scale=0.25]{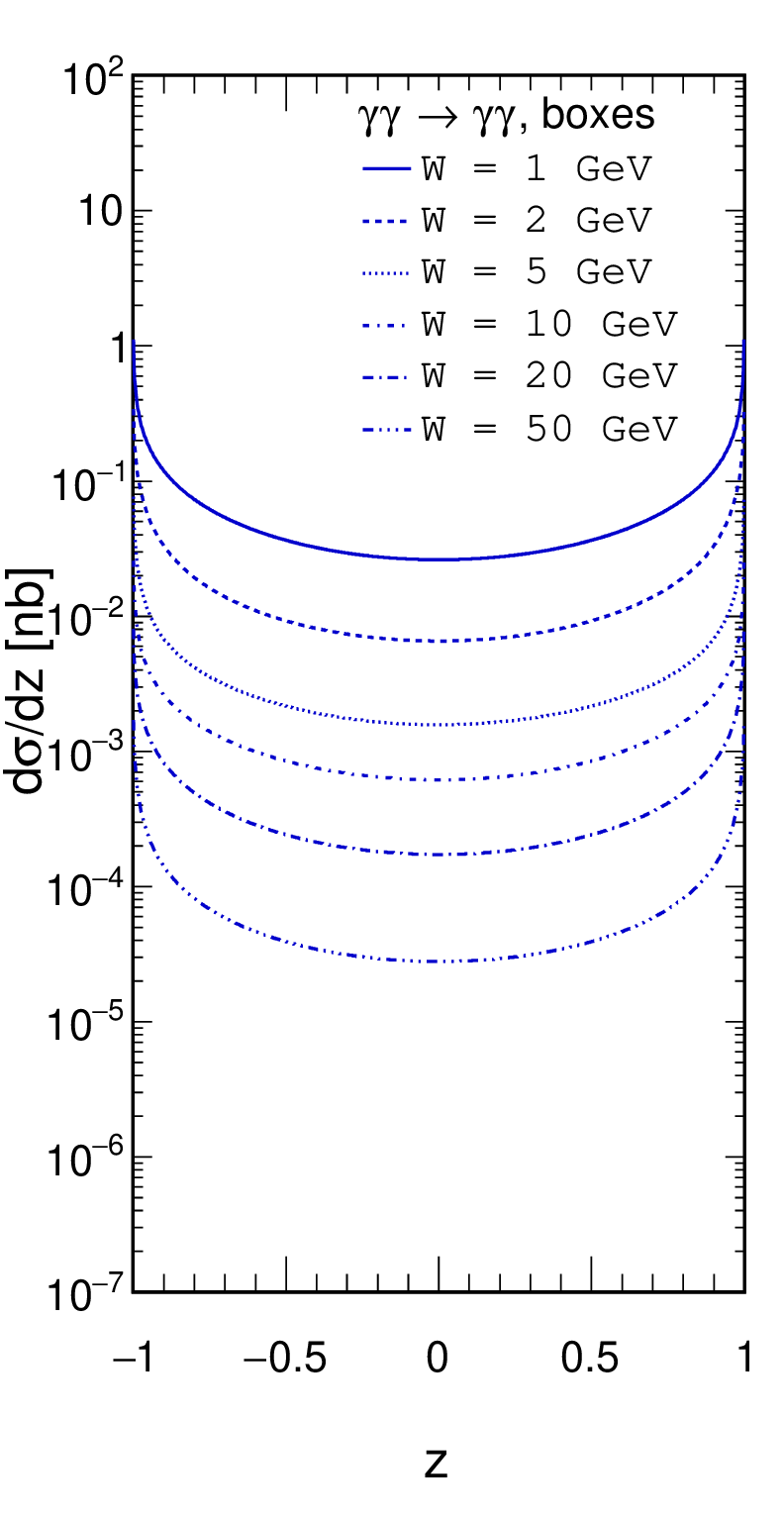}
	(b)\includegraphics[scale=0.25]{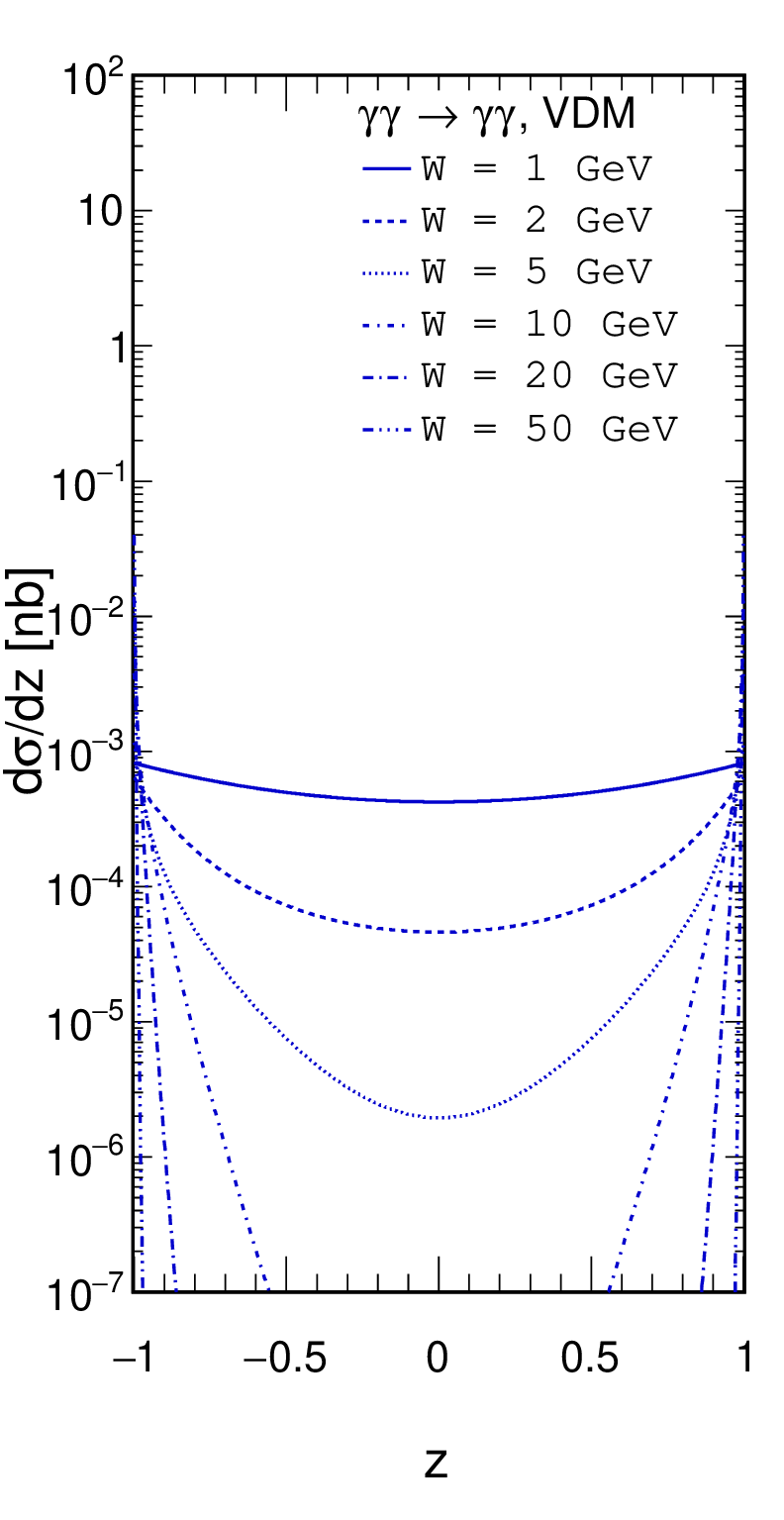}
	(c)\includegraphics[scale=0.25]{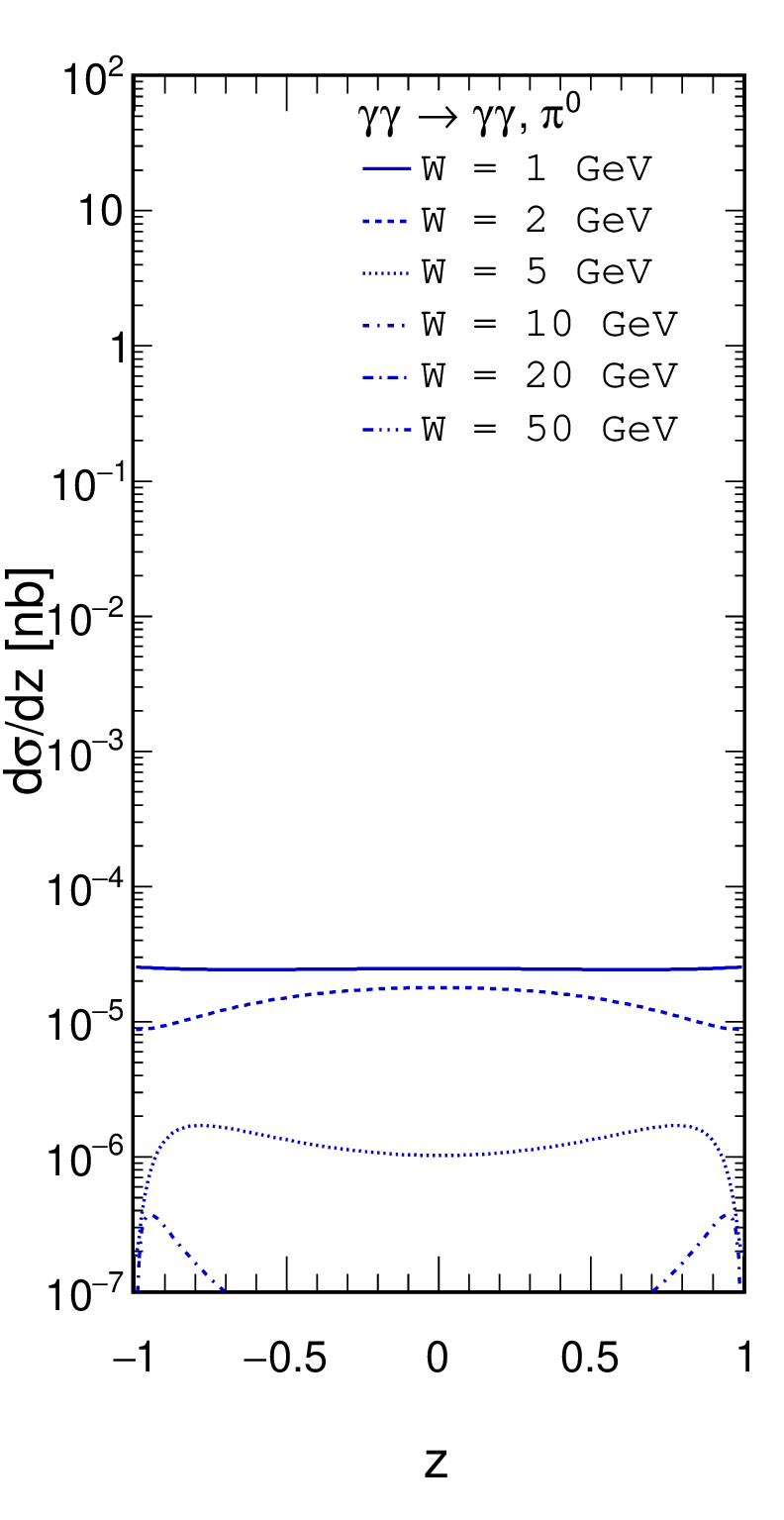}
        \end{center}
	\caption{$cos(\theta)$ distributions for (a) boxes, (b) double hadronic fluctuations
	and (c) $\pi^0$ exchange for different
	photon-photon collision energies $W= 1, 2, 5, 10, 20, 50$~GeV.
	}
	\label{fig:gamgam_gamgam_subleading}
\end{figure}

In Fig.~\ref{fig:gamgam_gamgam_subleading} we show $d \sigma / dz$ for 
$\gamma\gamma \to \gamma\gamma$
for (a) boxes, (b) double hadronic fluctuation calculated within 
the VDM-Regge approach
and (c) the $\pi^0$-exchange calculated as in Ref.~\cite{LS2017}. 
Results are presented for five energies in the range of $(1-50)$~GeV.
At larger energies, the VDM-Regge contribution peaks at $z = \pm 1$.
On the other hand, the $\pi^0$ exchange contribution has minima 
at $z = \pm 1$ which is due to the structure of corresponding
vertices. The latter contribution is relatively small. In general, the
box contributions dominate, especially for low photon-photon 
scattering energies. At larger scattering energies 
($W_{\gamma\gamma} >2$  GeV) the VDM-Regge contribution competes with 
the box contributions only at $z \sim \pm 1$.
Can one expect sizeable interference effects of both mechanisms?

\begin{figure}[!h]
(a)\includegraphics[scale=0.32]{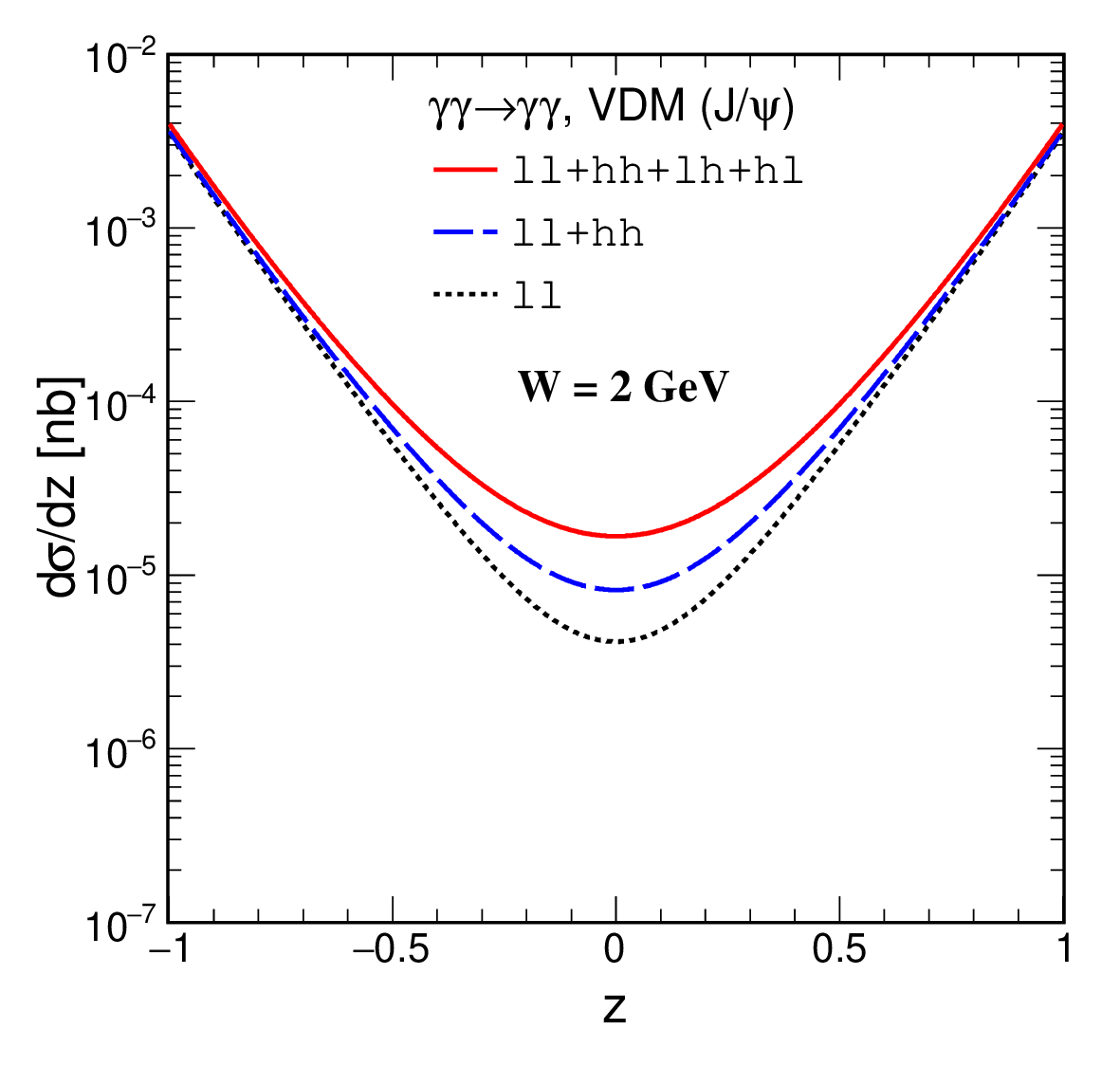}
(b)\includegraphics[scale=0.32]{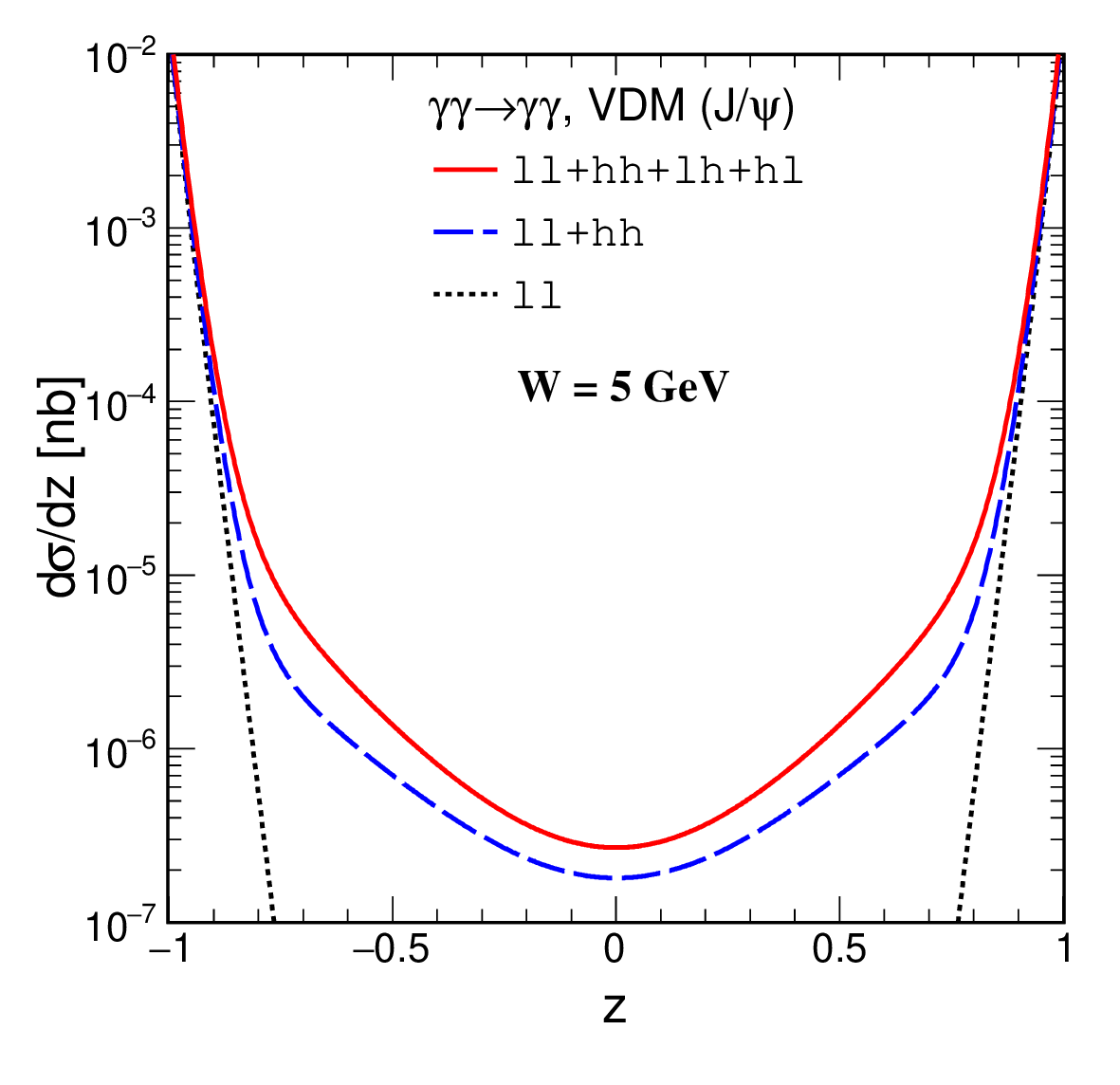}
\caption{Modification of $d\sigma/dz$ due to inclusion of
fluctuations of photons into virtual $J/\psi$ mesons: 
(a) $W = 2$~GeV, (b) $W = 5$~GeV.
The top solid line includes all components (light (\textit{l}) and heavy (\textit{h}) vector mesons), the dotted line only light vector mesons.
}
\label{fig:dsig_dz_vdm_hard} 
\end{figure}

Now we wish to discuss briefly the second biggest contribution:
double photon fluctuations.
The results are shown in Fig.\ref{fig:gamgam_gamgam_subleading}. 
In \cite{JKS2024} we includeed both light vector mesons $\rho^0$,
$\omega$, $\phi$) as well as $J/\psi$ (one or two) as described in the
theoretical section.
Our results, for two collision energies ($W = 2, 5$~GeV), are 
shown in Fig.\ref{fig:dsig_dz_vdm_hard}. The dotted line includes only light
vector meson fluctuations, the dashed line in addition double 
$J/\psi$ fluctuations and the solid line all combinations of photon 
fluctuations.
The inclusion of  $J/\psi$ meson fluctuations leads to an enhancement 
of the cross section at -0.5 $< z <$ 0.5. The enhancement is more 
spectacular for larger collision energy. The corresponding cross 
section there is, however, much smaller than the box contribution 
(see Fig.\ref{fig:gamgam_gamgam_subleading}).

\begin{figure}[!h]
        \begin{center}
	\includegraphics[scale=0.32]{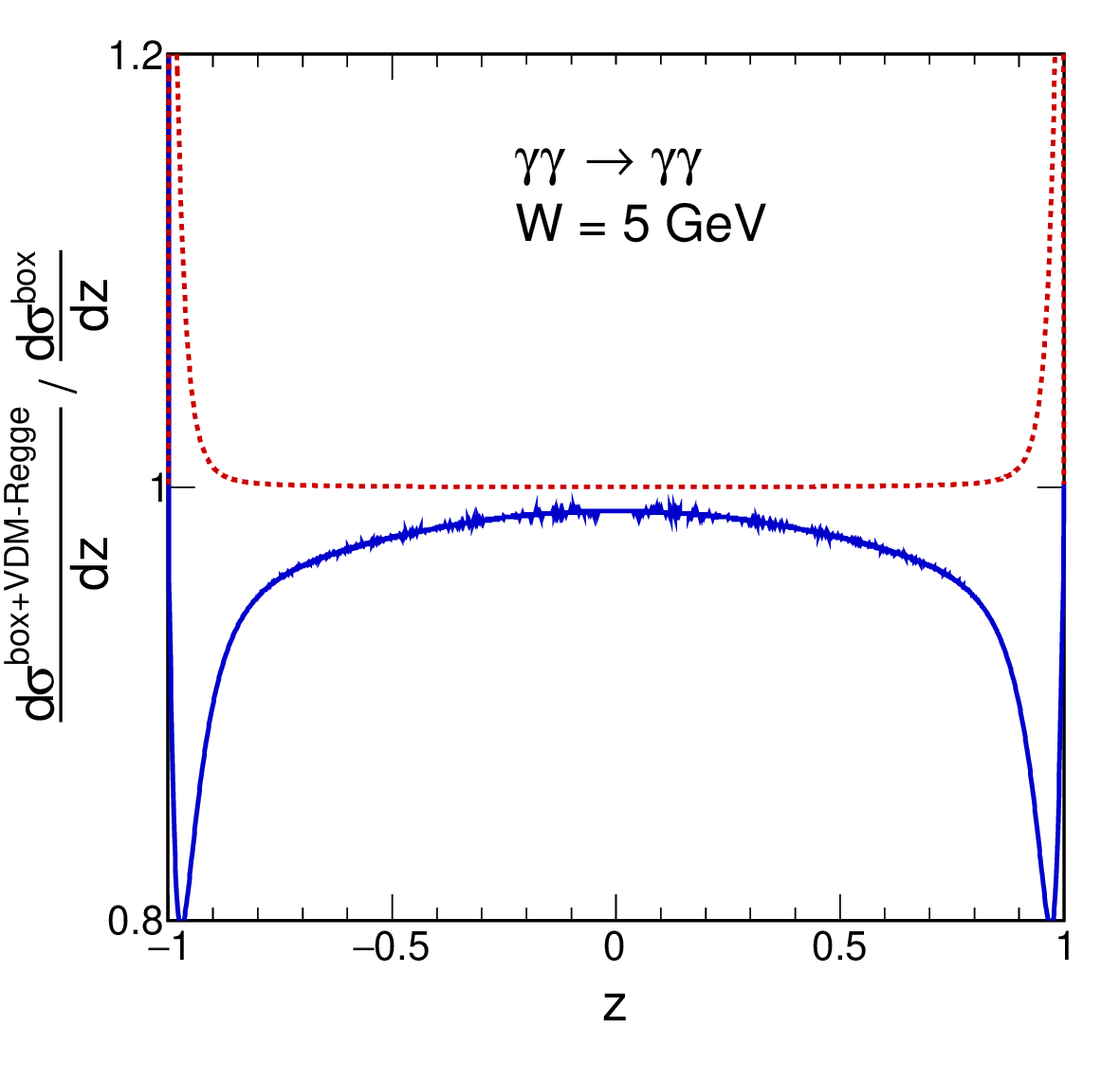}
        \end{center}
	\caption{The ratio of the coherent (blue) and incoherent (red) sum 
		of the box and VDM-Regge contributions divided by the cross section for the box contribution alone for $W = 5$~GeV.}
	\label{fig:ratio_z}
\end{figure}

Now we wish to concentrate on how the elementary cross section
changes when adding the box and VDM-Regge contributions.
This is shown in Fig.\ref{fig:ratio_z}.
The red line represents the incoherent sum,
while the blue line includes also interference effects.
In this calculation, the so-called ``sqrt'' trajectories 
\cite{Brisudova1,Brisudova2} were used.
one observes a negative interference effect. Adding the remaining
contributions would lead to additional deviations.

\subsection{Heavy ion UPC}

\begin{figure}[!h]
	\begin{minipage}{0.48\columnwidth}
		(a)\includegraphics[scale=0.3]{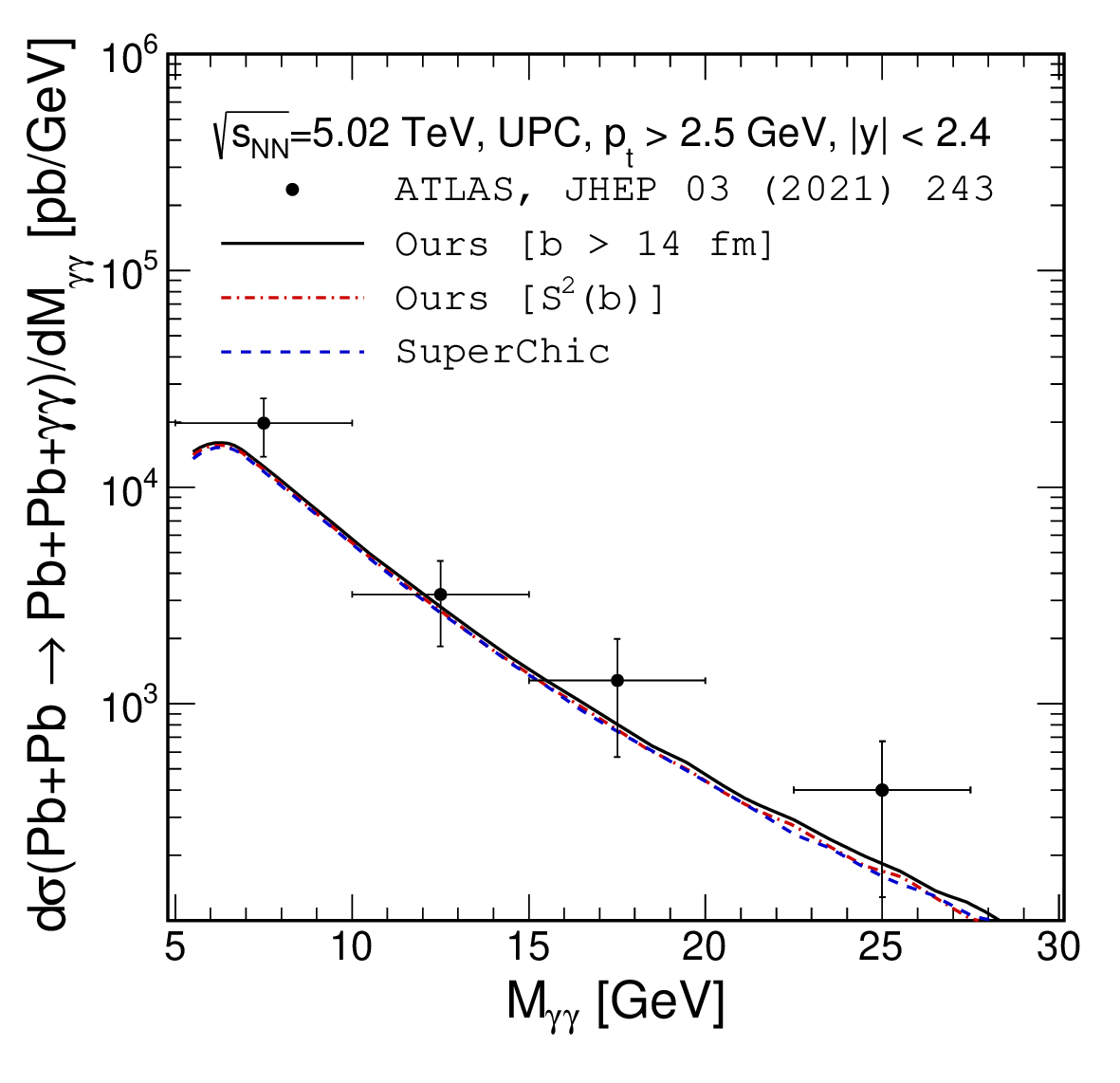}
	\end{minipage}
	\begin{minipage}{0.48\columnwidth}
		(b)\includegraphics[scale=0.3]{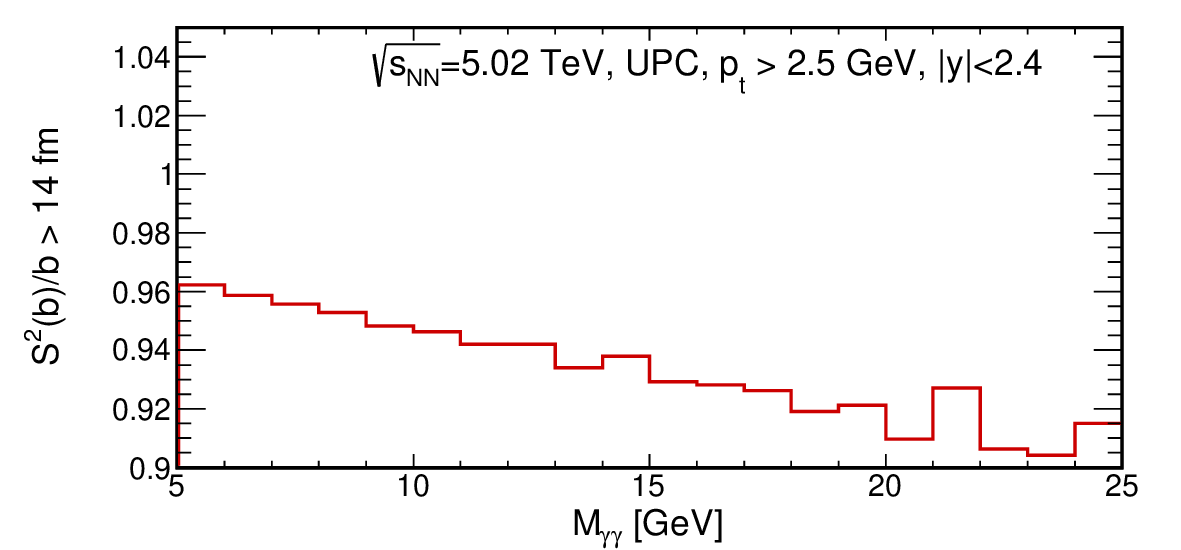}\\
		(c)\includegraphics[scale=0.3]{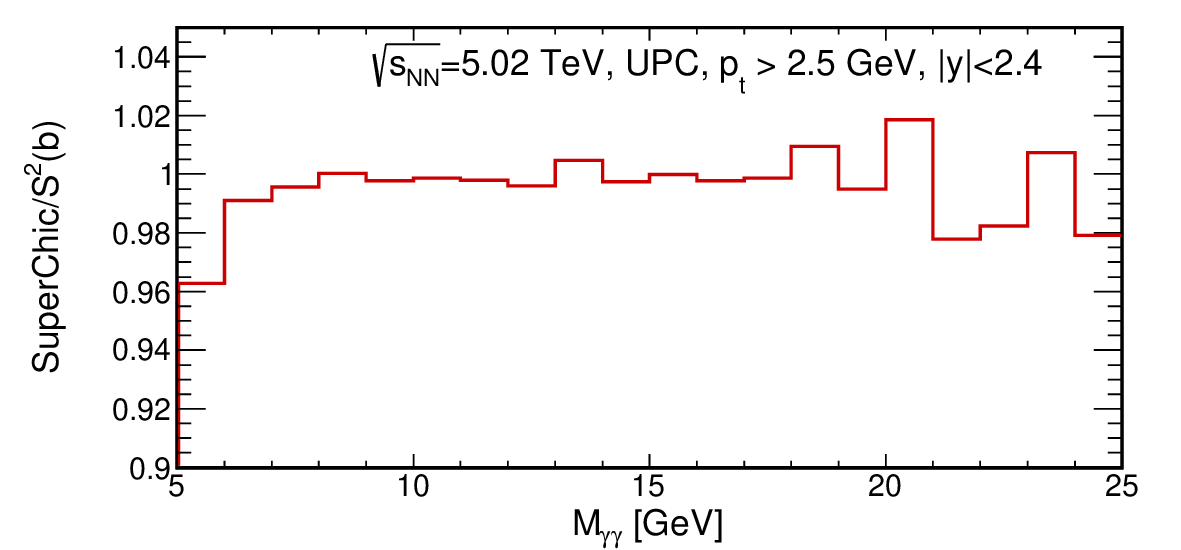}
	\end{minipage}
\label{fig:2}
	\caption{Differential cross section as a function
          of two-photon invariant mass at $\sqrt{s_{NN}}$ = 5.02 $~$ 
          TeV. 
          (a)
          The ATLAS experimental data are collected together with 
          theoretical
          results including a sharp cut on impact parameter 
          ($b >$ 14 fm -
          solid black line) and smooth nuclear absorption factor
          $S^2(b)$ (dash-dotted red line). For completeness, results
          that are obtained with the help of Eq.~(\ref{eq:tot_xsec}) 
          are compared with results from SuperChic.
         The right panel shows two ratios:
          (b) ratio of distributions calculated by us with sharp and 
 smooth cut-off on impact parameter and (c) the ratio of SuperChic 
 result to our result, using a smooth representation of the gap 
 survival factor.   
}
\end{figure}

To summarize the present status of $\gamma \gamma \to \gamma \gamma$
scattering in Fig.\ref{fig:2} we confront results of our calculation 
with current ATLAS data \cite{ATLAS2}.
We discuss also how the results depend on the treatment of absorption
corrections.
The results of the two different approximations 
(as described in the figure caption) almost coincide.
For comparison, we show also results obtained with the SuperChic 
generator \cite{HarlandLang}.
In general, there is reasonable agreement of the Standard Model
predictions with the ATLAS data. Similar agreement is achieved for
the CMS data.

\begin{figure}[!h]
\begin{center}
\includegraphics[width=5cm]{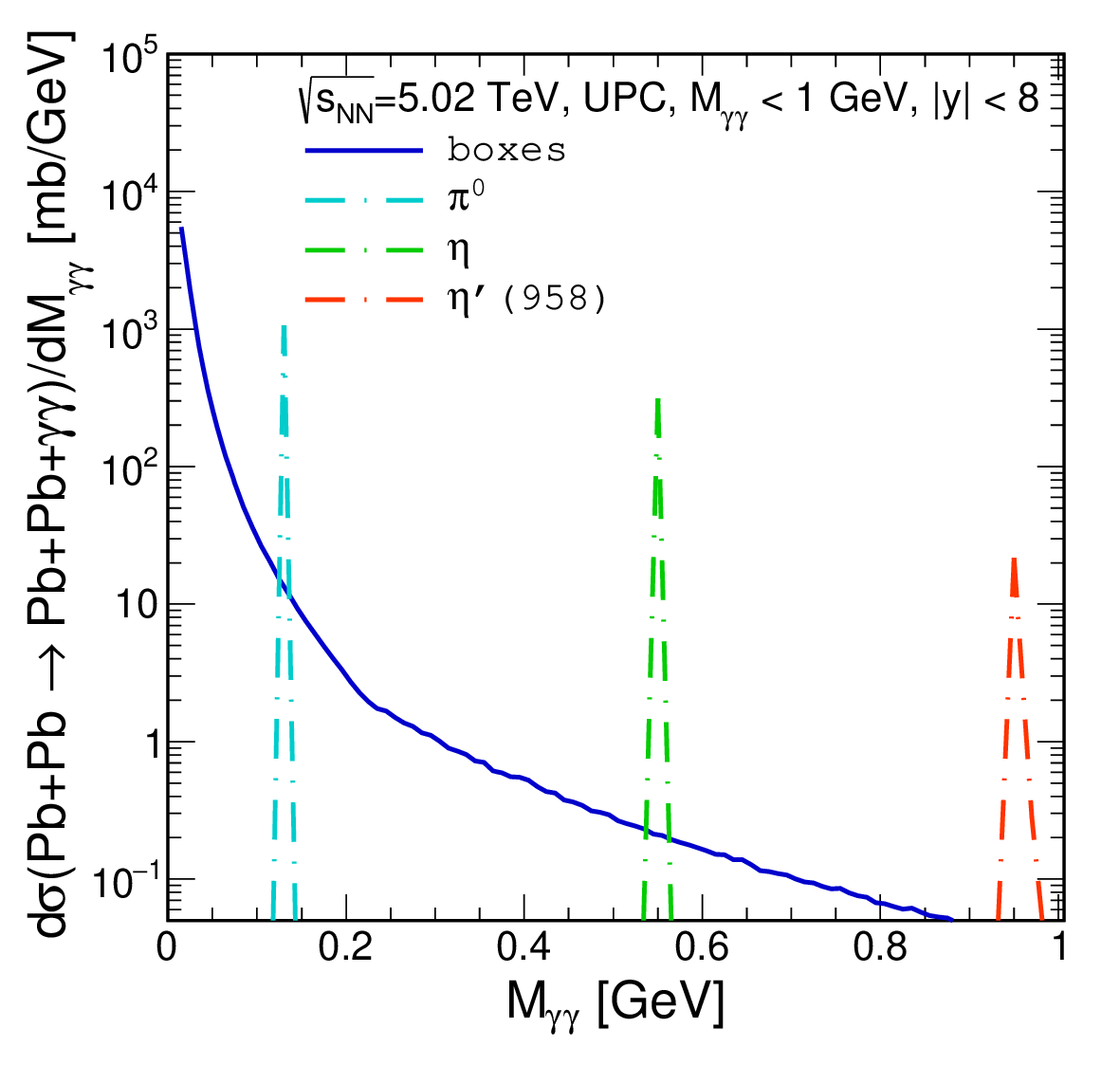}
\includegraphics[width=5cm]{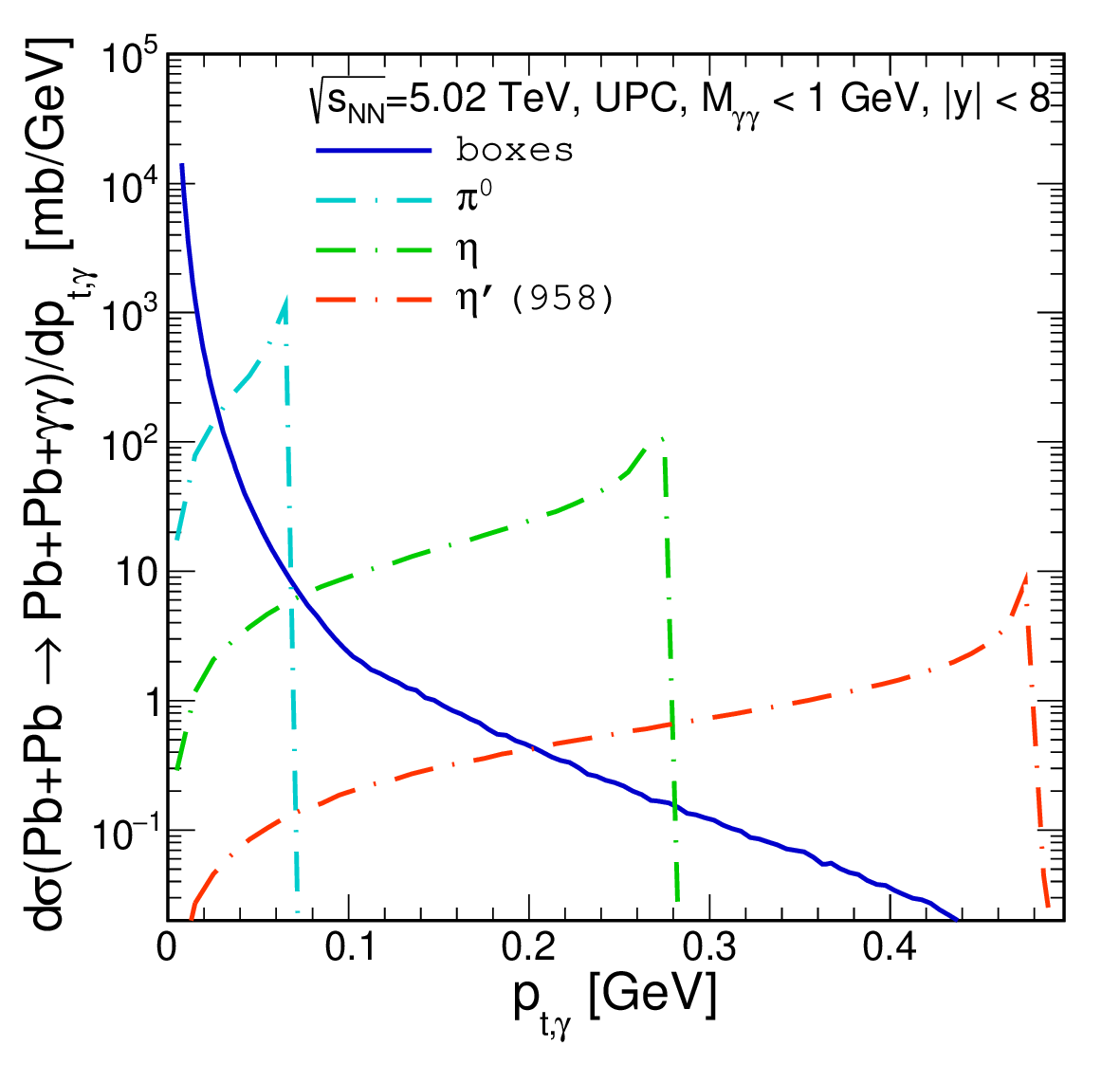}
\end{center}
\caption{\label{fig:resonances}
Box versus resonance contributions for diphoton invariant mass (left)
and photon transverse momentum (right).}
\end{figure}

Many light vector mesons have large coupling to two photons.
In \cite{LS2017} for a first time the role of resonances was 
discussed, for elementary cross section only.
In Fig.\ref{fig:resonances} we show the contributions of light
mesonic photon-photon resonances for heavy ion UPC. The results shown
are for a broad range of photon rapidities and transverse momenta.
We show distribution in diphoton inariant mass (left) and photon
transverse momentum (right). It is obvious that elimination
of the resonance contributions may be difficult. But perhaps it
is not necessary as the mesonic resonance contributions are a part
of photon-photon scattering. The cuts used by ATLAS or CMS
allowed to eliminate the contributions of light mesons.
The contribution of heavy mesons is expected to be small.

\begin{figure}[!h]
        \begin{center}
	(a)\includegraphics[scale=0.32]{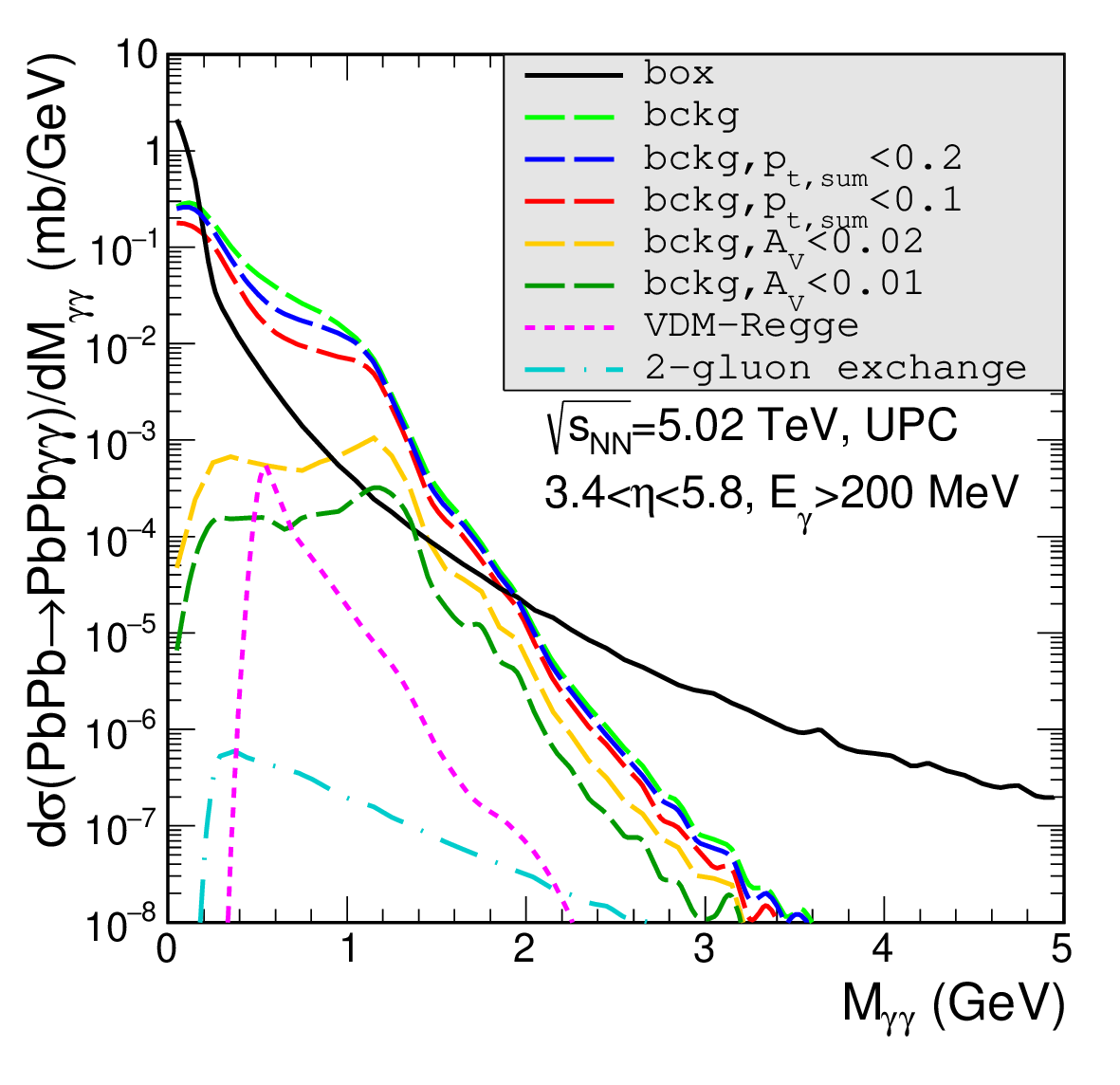}
	(b)\includegraphics[scale=0.32]{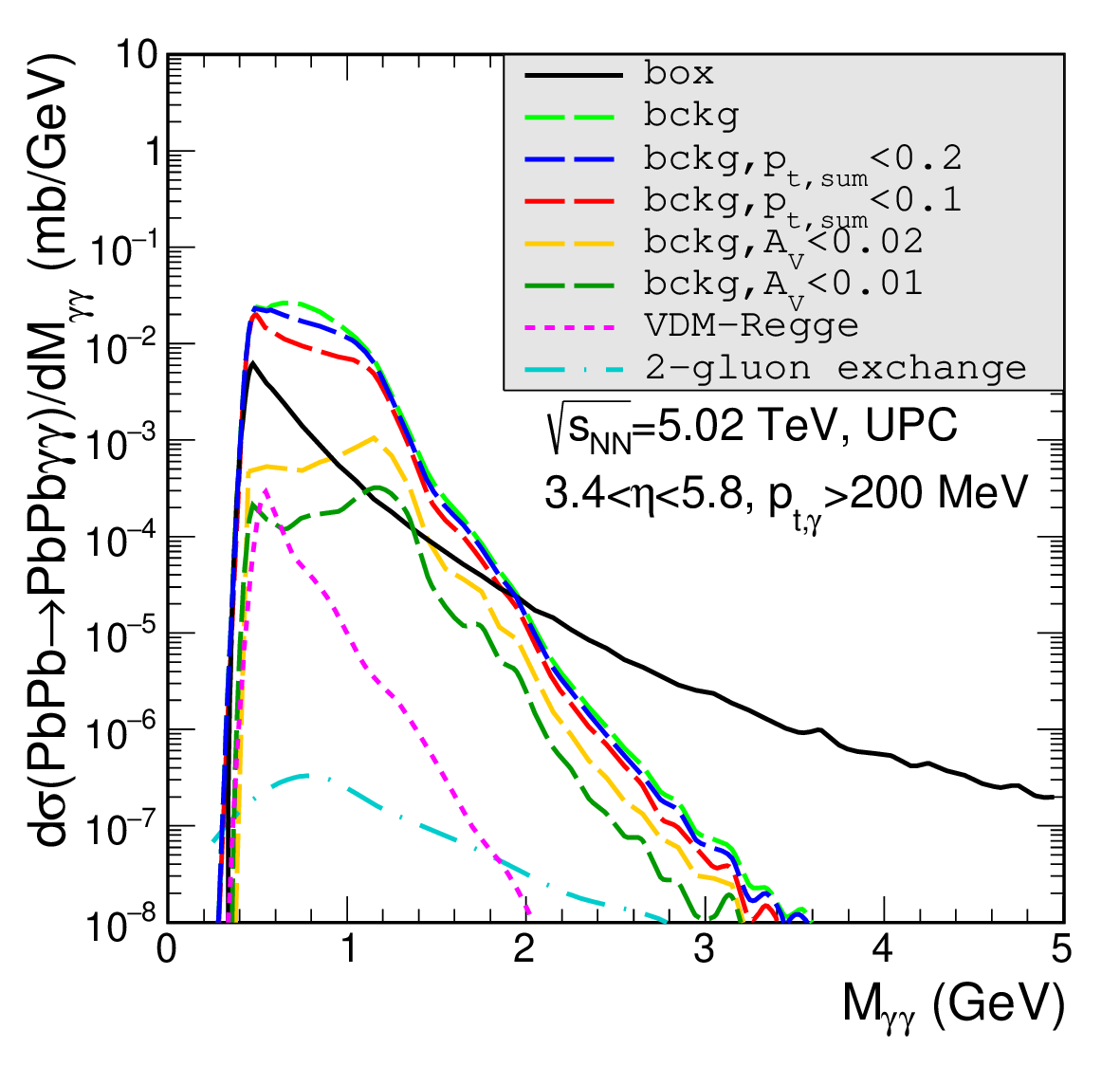}
        \end{center}
	\caption{\label{fig:dsig_dW_ALICE_FoCal_2} Diphoton 
invariant mass distribution for the UPCs. Predictions are made for the future FoCal acceptance, i.e. (a) $E_{t,\gamma}>200$~MeV and $3.4<y_{\gamma_{1/2}}<5.8$, (b) $p_{t,\gamma}>200$~MeV and $3.4<y_{\gamma_{1/2}}<5.8$. Here, both photons are ''measured'' in FoCal. The background contribution is presented for different cuts.}
\end{figure}

The FoCal detector planned for Run 4 was described in \cite{FoCal}. 
It is a general purpose detector. It can also measure photons.
We start our presentation from the results when both 
photons are measured by FoCal. In 
Fig.\ref{fig:dsig_dW_ALICE_FoCal_2} 
(a) we show results when both photons have energies bigger than
200~MeV. In addition, we show the contribution of the $\pi^0\pi^0$ 
background. 
In this case, only two photons are measured. Without additional cuts, 
the background is clearly bigger than the signal. However, by 
imposing extra conditions on vector asymmetry, we can lower the 
background contribution.

\begin{figure}[!h]
	(a)\includegraphics[scale=0.3]{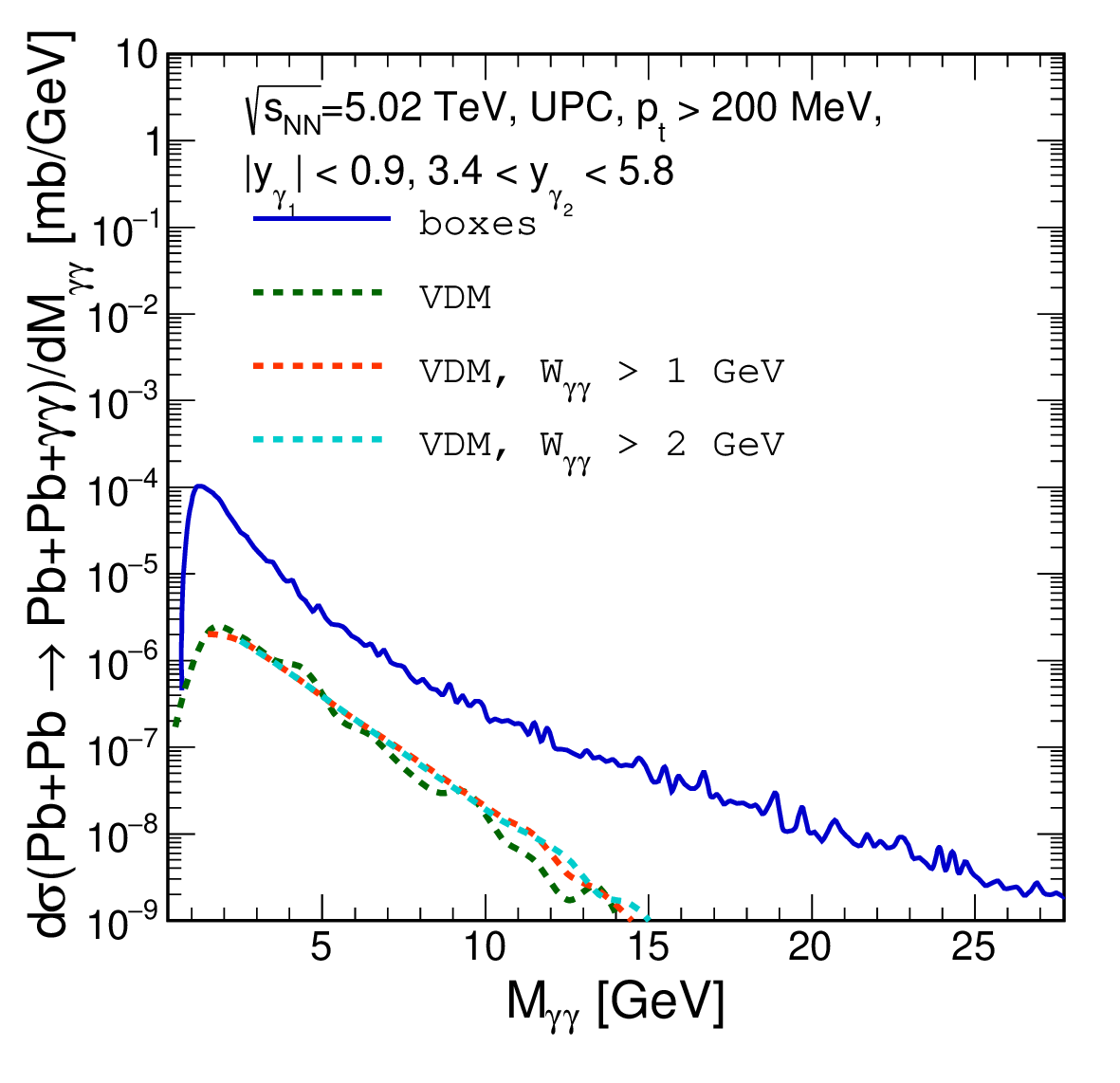}
	(b)\includegraphics[scale=0.3]{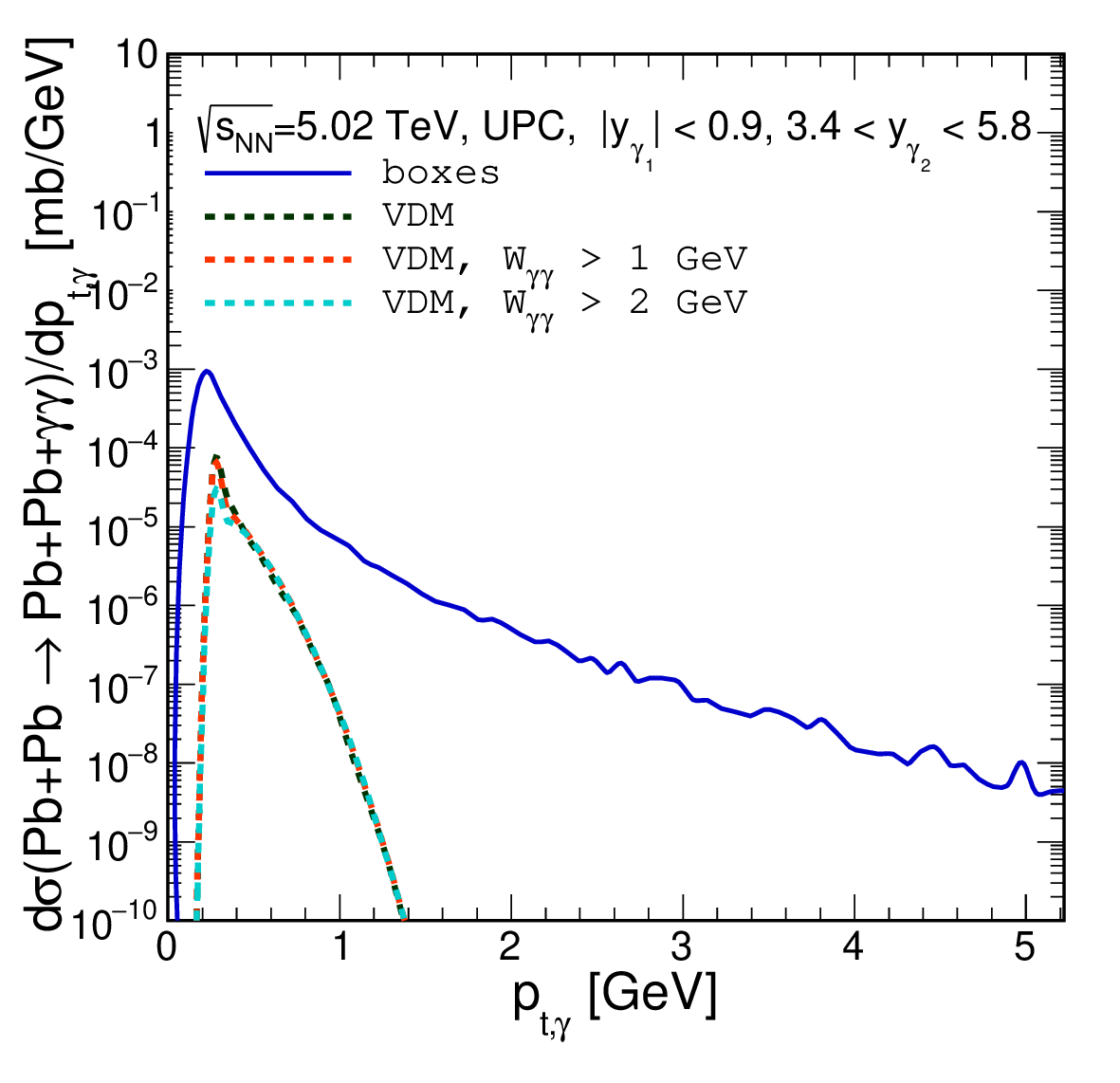}
    (c)\includegraphics[scale=0.5]{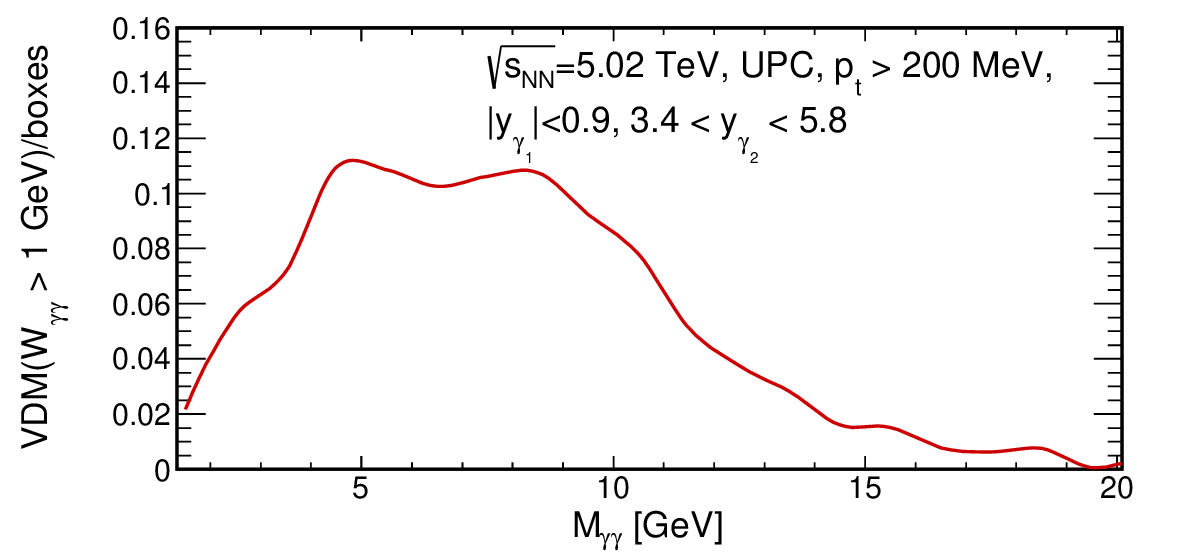}
	\caption{\label{fig:15} Prediction for the FoCal detector in 
association with mid-rapidity ALICE detector for photons: 
$p_{t}>200$~MeV, diphoton mass $M_{\gamma\gamma}$ $>$ 400~MeV 
and photon rapidities $|y_{1}|< 0.9$ and $y_{2} \in (3.4,5.8)$. 
The blue line corresponds to fermionic loops and the green lines 
to the VDM-Regge contribution. (a) Diphoton invariant mass 
distribution, (b) photon transverse momentum distribution, 
(c) ratio of the VDM-Regge ($W_{\gamma\gamma}$ $>$ 1~GeV) and
  box contributions as a function of di-photon invariant mass.
No interference effects were included here.}
\end{figure}

In Fig.\ref{fig:15} we show similar distributions but for
$p_t$~$>$~0.2~GeV and combined ALICE and FoCal rapidity region. 
Here in some regions of the phase space, the VDM-Regge contribution
could be seen as $~10\%$ modification of the cross section with respect 
to the calculations with only boxes.
Here the separated VDM-Regge component is even bigger. We conclude that
already at Run 4 one could indirectly observe a signature of
mechanisms other than fermionic boxes.

Now we will show distributions relevant for the ALICE 3 detector.

\begin{figure}[!h]
        \begin{center}
	\includegraphics[scale=0.32]{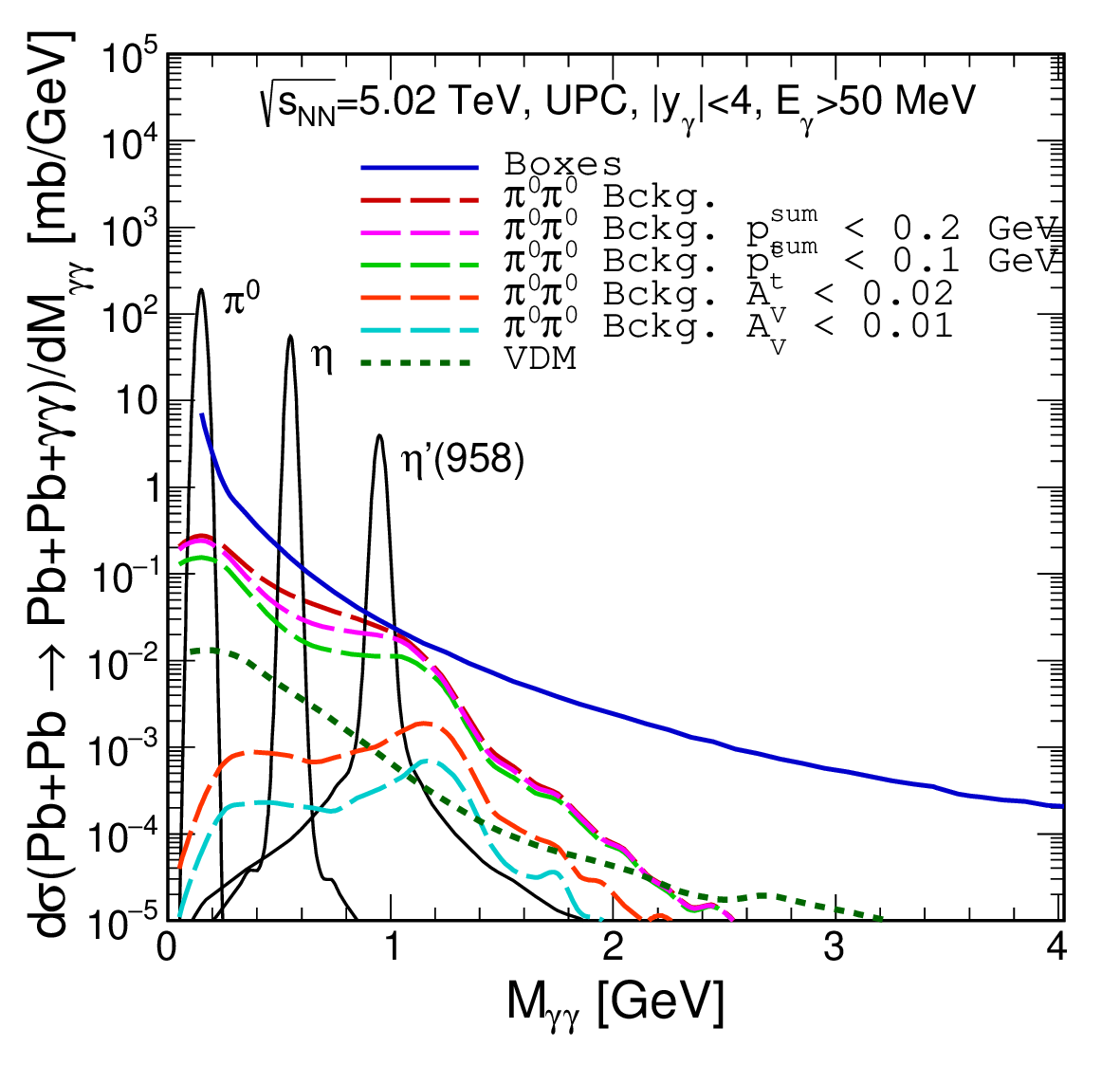}
        \end{center}
	\caption{\label{fig:ALICE3_m} Diphoton invariant mass 
distribution for ALICE 3, i.e. rapidity $y_{1/2} \in (-4,4)$ 
and photon energy E$_{\gamma}>50$~MeV. Here the blue solid line 
relates to the box contribution, the dotted line to the VDM-Regge 
component and the dashed lines are for double-$\pi^0$ background 
contribution. Here we impose several extra conditions on diphoton 
transverse momenta and vector asymmetry.}
\end{figure}

In Fig.~\ref{fig:ALICE3_m} we show distributions in diphoton
invariant mass for photons $-4< y_1, y_2 <4$ and $E_{\gamma} >50$~MeV 
(see Ref.~\cite{ALICE3}). We show the light-by-light box
contribution (solid line) as well as the $\pi^0 \pi^0$ background 
contribution (red lines).
At diphoton invariant masses, $0.5$~GeV~$<M_{\gamma\gamma}<1$~GeV, 
the background contribution is almost as big as the signal contribution. 
As discussed in \cite{KNSS}
it can be to some extent reduced. Although the background is smaller
than fermionic boxes in the full range of diphoton invariant mass, it
can be further reduced by imposing the cut on $|\vec{p_{1t}} +
\vec{p_{2t}}|<0.1$~GeV and vector asymmetry $A_V<0.02$. Imposing a cut
on the background causes that the background in the whole 
diphoton invariant mass range is much smaller than the signal.

\begin{figure}[!h]
        \begin{center}
	\includegraphics[scale=0.32]{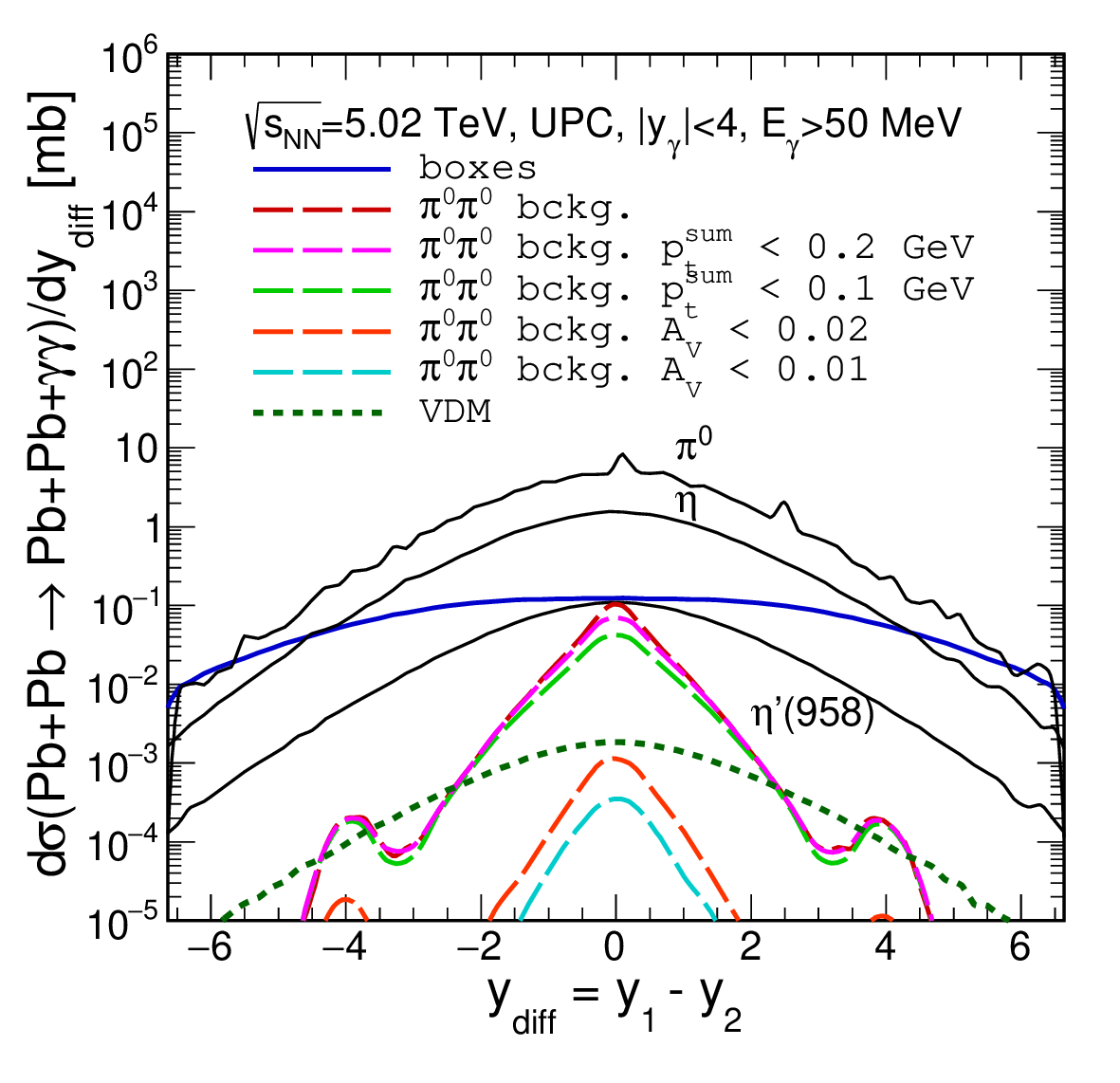}
        \end{center}
	\caption{\label{fig:ALICE3_ydiff} Differential cross section as
          a function of $y_{diff} = y_1 - y_2$ for extended ALICE 3
          kinematics: 
$|y_{1/2}|<4$ and $E_\gamma>50$~MeV. Results are presented for boxes, 
resonances, VDM-Regge and double-$\pi^0$ background.}
\end{figure}

In Fig.~\ref{fig:ALICE3_ydiff} we show distribution in
$y_{diff}=y_1-y_2$. Again different contributions are shown
separately. The results for the double-$\pi^0$ background contribution
are particularly interesting. It has a maximal contribution at
$y_{diff}=0$ and drops quickly for larger $|y_{diff}|$. An extra cut on
$y_{diff}$ could therefore considerably reduce the unwanted
double-$\pi^0$ contribution. In Fig.~\ref{fig:ALICE3_ydiff} 
we show what happens when imposing the cut on $y_{diff}$. 
The effect of such a cut on box contribution is relatively small 
but leads to a huge reduction of the $\pi^0 \pi^0$ background. 
The effect of the cut is much larger for small $M_{\gamma\gamma}$ 
and therefore could be avoided if one is interested in 
this region of energies.

\begin{figure}[!h]
	(a)\includegraphics[scale=0.32]{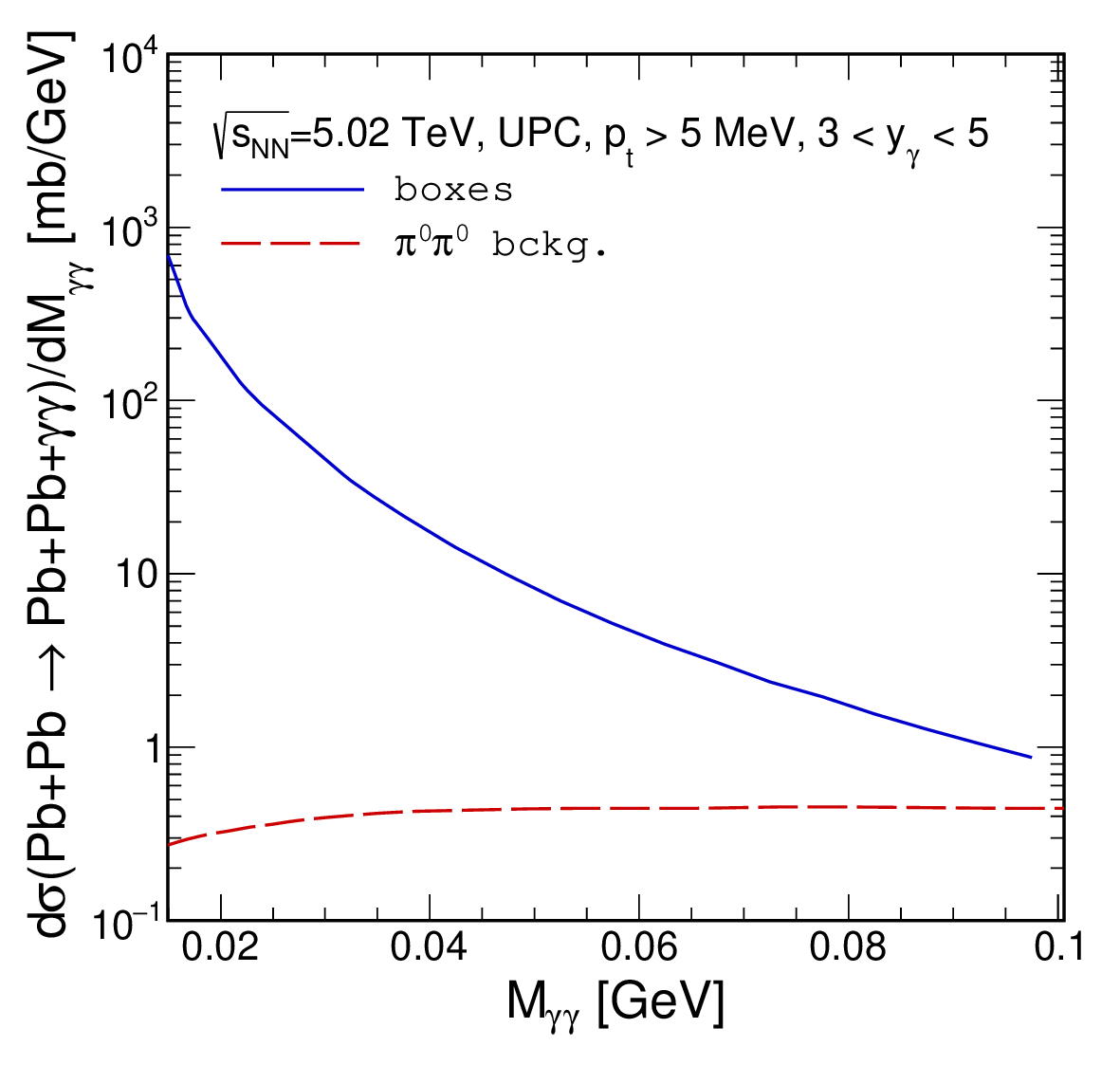}
	(b)\includegraphics[scale=0.32]{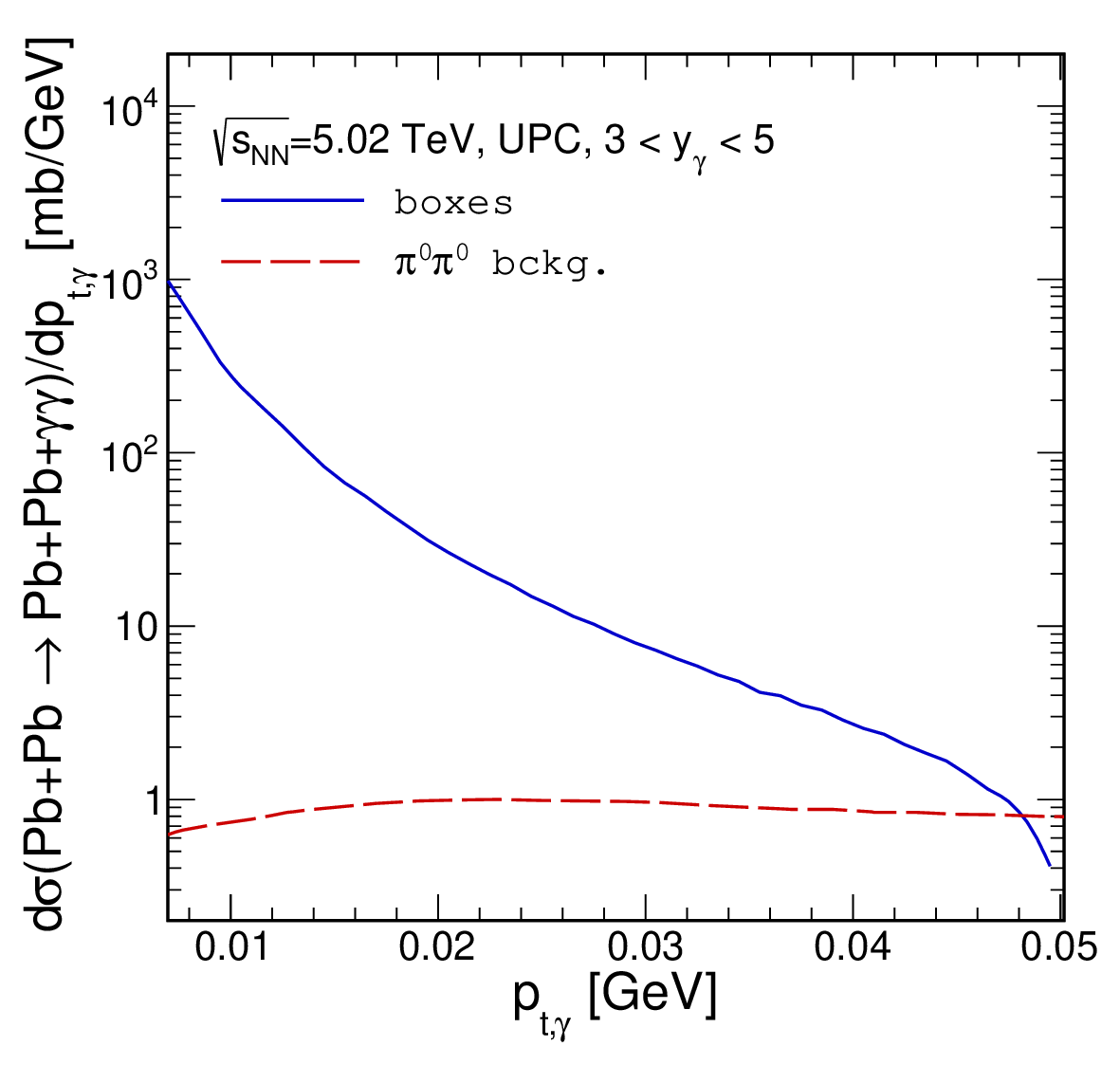}
	\caption{\label{fig:13} Prediction for the ALICE 3 
experiment for soft photons: $p_{t} = (5-50)$~MeV and photon 
rapidities $y_i \in (3,5)$ . The blue line corresponds to 
fermionic loops, the red line relates to the double-$\pi^0$ 
background. 
(a) diphoton invariant mass distribution, 
(b) photon transverse momentum distribution.}
\end{figure}

In Fig.\ref{fig:13} we show distribution in $M_{\gamma \gamma}$ 
(a) and $p_t$ (b) for a planned special photon detector in forward 
direction $3< y_\gamma <5$. Here $p_t >5$~MeV was imposed as 
described in Ref.~\cite{ALICE3}. 
We show that at low $M_{\gamma\gamma}$ and low $p_t$ 
the LbL signal by far exceeds the $\pi^0 \pi^0$ background, 
even without including any suppression condition for the
background. 
Here we assumed 2 $\pi$ azimuthal coverage of the
planned special photon detector.
The special detector having the planned very small transverse momentum
coverage should allow to measure completely new,
low energy, regime of photon-photon scattering.
Here both background and mesonic resonance contributions should
be absent.

\section{Conclusions}

Recently we discussed different mechanisms of 
$\gamma \gamma \to \gamma \gamma$
scattering such as leptonic/quarkish boxes, double hadronic
fluctuations, neutral $t/u$-channel pion exchanges and two-gluon
 exchanges. Possible effects of the searched for subleading mechanisms 
have been discussed. The latter contributions turned out
difficult to be identified in  ATLAS and CMS measurements.
We have discussed possible interference effect of box and
double-hadronic fluctuations for $\gamma\gamma \to \gamma\gamma$ 
scattering for future measurements.

In the literature only the box contributions were discussed before.
We have tried to identify the region where the other contributions
could appear.
In addition we discussed how to reduce the 
unwanted $\gamma \gamma \to \pi^0 \pi^0$ background.

The FoCal project does not seem to allow for breakthroughs
for LbL scattering. but may be used to supplement the ALICE,
not yet officially presented, experimental studies.

We have also made predictions for the ALICE 3 ($-4<y_\gamma<4$) 
and for a planned special soft photon detector ($3<y_\gamma<5$). 
We have shown that by imposing a cut on $y_{diff}=y_1-y_2$ one 
can efficiently eliminate the unwanted $\pi^0 \pi^0$ background. 
The soft photon detector can be used to measure the 
$\gamma\gamma\to\gamma\gamma$ scattering at extremely small energies, 
$W_{\gamma\gamma}<0.05$~GeV.  
Therefore we conclude that the ALICE 3 infrastructure will be 
extremely useful to study the $\gamma \gamma \to \gamma \gamma$ 
scattering in a new, not yet explored, domain of energies and 
transverse momenta. In this domain the $\pi^0 \pi^0$ background 
can to large extent be eliminated.

In our recent calculations we used EPA in the impact parameter space.
In the future one can try to use also so-called Wigner function 
approach used for $e^+ e^-$ production in semi-peripheral
lead-lead collisions (never used for the di-photon production).
This goes, however, beyond the scope of the first exploratory study.

\section{Acknowledgments}

This work was partially supported by the Polish National Science Center
grant UMO-2018/31/B/ST2/03537 and by the Center for Innovation and
Transfer of Natural Sciences and Engineering Knowledge in Rzesz\'ow.






\section*{References}

\end{document}